\documentclass[12pt,preprint]{aastex}
\pdfoutput=1
\usepackage{emulateapj5}
\usepackage{natbib}
\usepackage{onecolfloat}
\usepackage{subfigure}  
\usepackage{color}

\def\aap{A\&A} 
\def\apj{ApJ} 
 
\def\apjl{ApJL} 
\def\mnras{MNRAS} 
\def\araa{ARA\&A}

\def\nat{Nature}

\def\apjs{ApJS} 

\def\gca{Geochimica et Cosmochimica Acta}

\newcommand{\minv}{$h^{-1} M_{\odot}$}

\newcommand{\bdm}{\begin{displaymath}} 
\newcommand{\edm}{\end{displaymath}}
\newcommand{\beq}{\begin{equation}} 
\newcommand{\eeq}{\end{equation}} 
\newcommand{\beqnarr}{\begin{eqnarray}}
\newcommand{\eeqnarr}{\end{eqnarray}}
\newcommand{\bit}{\begin{itemize}} 
\newcommand{\eit}{\end{itemize}} 
\newcommand{\ben}{\begin{enumerate}} 
\newcommand{\een}{\end{enumerate}}
\newcommand{\bfi}{\begin{figure}[htb]} 
\newcommand{\bpfi}{\begin{figure}[p]}
\newcommand{\barr}{\begin{array}}
\newcommand{\earr}{\end{array}}
\newcommand{\bec}{\begin{center}}
\newcommand{\eec}{\end{center}}
\newcommand{\bs}{\begin{sideways}}
\newcommand{\es}{\end{sideways}}

\shorttitle{Simulated CC and NCC}
\shortauthors{Rasia et al.}
\begin{document}
\twocolumn[%
\title{Cool Core Clusters from Cosmological Simulations}
\author{
E. Rasia\altaffilmark{1,2},
S. Borgani\altaffilmark{1,3,4},
G. Murante\altaffilmark{1},
S. Planelles\altaffilmark{1,3,4}
A. M. Beck\altaffilmark{5},
V. Biffi\altaffilmark{1,3},
C. Ragone-Figueroa\altaffilmark{6},
G. L. Granato\altaffilmark{1},
L. K. Steinborn\altaffilmark{5},
K. Dolag\altaffilmark{5,7}
}
\affil{$^1$ INAF, Osservatorio Astronomico di Trieste, via Tiepolo 11,
  I-34131, Trieste, Italy, rasia@oats.inaf.it} 
\affil{$^2$ Department of Physics, University of Michigan, 450 Church St., Ann Arbor, MI  48109, USA}
\affil{$^3$ Dipartimento di Fisica dell' Universit\`a di Trieste,
  Sez. di Astronomia, via Tiepolo 11, I-34131 Trieste, Italy} 
\affil{$^4$ INFN, Instituto Nazionale di Fisica Nucleare, Trieste, Italy}
\affil{$^5$ Universit\"ats-Sternwarte M\"unchen, Scheinerstr.1, D-81679 M\"unchen, Germany}
\affil{$^6$Instituto de Astronom\'a Te\'orica y Experimental (IATE),
  Consejo Nacional de Investigaciones Cient\'ificas y T\'ecnicas de la
  Rep\'ublica Argentina (CONICET), Observatorio Astron\'omico,
  Universidad Nacional de C\'ordoba, Laprida 854, X5000BGR, C\'ordoba,
  Argentina} 
\affil{$^7$ Max-Plank-Institut f\"ur Astrophysik, Karl-Schwarzschild
  Strasse 1, D-85740 Garching, Germany}

\begin{abstract}

  We present results obtained from a set of cosmological hydrodynamic
  simulations of galaxy clusters, aimed at comparing predictions with
  observational data on the diversity between cool-core (CC) and
  non-cool-core (NCC) clusters. Our simulations include the effects of
  stellar and AGN feedback and are based on an improved version of the
  smoothed particle hydrodynamics code GADGET-3, which ameliorates gas
  mixing and better captures gas-dynamical instabilities by including
  a suitable artificial thermal diffusion.  In this {\em Letter}, we
  focus our analysis on the entropy profiles, the primary diagnostic we used
  to classify the degree of cool-coreness of clusters, and on the iron
  profiles. In keeping with observations, our simulated clusters
  display a variety of behaviors in entropy profiles: they range from
  steadily decreasing profiles at small radii, characteristic of
  cool-core systems, to nearly flat core isentropic profiles,
  characteristic of non-cool-core systems. Using observational
  criteria to distinguish between the two classes of objects, we find
  that they occur in similar proportions in both simulations and in
  observations. Furthermore, we also find that simulated cool-core
  clusters have profiles of iron abundance that are steeper than those
  of NCC clusters, which is also in agreement with observational
  results.  We show that the capability of our simulations to generate
  a realistic cool-core structure in the cluster population is due to
  AGN feedback and artificial thermal diffusion: their combined action
  allows us to naturally distribute the energy extracted from
  super-massive black holes and to compensate for the radiative losses of
  low-entropy gas with short cooling time residing in the cluster
  core.
\end{abstract}
]

\section{Introduction}\label{sec:intro} 

During their hierarchical assembly, galaxy clusters grow in mass via
diffuse accretion processes as well as via 
merger events. These processes lead to the shock heating of the intra-cluster medium
(ICM) to a virial temperature of up to $T\sim 10^8$ K, with central
electron number densities corresponding to $n_{\rm e}\sim10^{-3}$cm$^{-3}$
\citep[e.g.][for a review]{voit05,kravtsov_borgani12}. At this
density, hot baryons in the core regions have a cooling time,
$t_{\rm cool} \propto T^{1/2} /n_{\rm e}$, shorter than the Hubble time. Under
this condition, the ICM suffers from radiative losses and becomes
colder and denser. This process might initiate a run-away cooling
process, which would cause an extreme accretion rate in the central
galaxy, unless some heating mechanism were to balance the
cooling. Substantial evidences from X-ray observations proved that the
regulating mechanism is the feedback by active galactic nuclei
\citep[AGNs, e.g.][]{mcnamara_nulsen07,voit.etal.2015}.

In a situation of steady accretion, low-entropy\footnote{We adopt the standard in the X--ray 
definition of entropy:
  $K=T/n_e^{2/3}$ with $T$ being the ICM temperature and $n_{\rm e}$ being the
  electron number density.}
  and metal-enriched gas associated with
merging substructures, or funneled by filaments, sink toward the central regions
of the so-called cool-core clusters (CC,
named as such by \citealt{molendi&pizzolato}).
%
%
On the
contrary, non-cool-core (NCC) objects are observed with nearly
isentropic gas cores at a higher entropy level and show no evidence of
a spike in their metal abundances \citep{degrandi.etal.2004,maughan.etal.2008}.

While idealized simulations reproduce some of the many features of the
CC systems \citep[e.g][]{gaspari.etal.2015,li.etal.2015}, producing
the diverse populations of CC and NCC clusters has so far proved to be so far
quite a formidable challenge for cosmological hydrodynamic simulations
\citep[][and references therein]{borgani&kravtsov}.  Earlier
simulation works predicted that the origin of the two classes of CC
and NCC clusters is ``primeval", with CC being destroyed exclusively
by early mergers or preheating phenomena
\citep{burns.etal.2008,mccarthy.etal.2008,poole.etal.2008,
  planelles_quilis}. In fact, a CC cluster was claimed to maintain its
high-density and metal-rich core even after a major merger, if that
happens at $z\leq 0.5$. However, this prediction is in conflict with
observations and recent cosmological simulations by 
\cite{hahn.etal.2015}. In most cases, CC clusters show regular X-ray morphology
while the opposite is true for NCC systems. 
Data, thus, favor an
``evolutionary" model in which a transition between the CC and NCC status
of a cluster can take place over relatively short timescales in
consequence of a merger event that mixes the convective stable gas and
destroys the CC \citep{rossetti.etal.2011}.
Specifically, \cite{hahn.etal.2015} advocate that the dichotomy 
between CC and NCC clusters 
originates by low-angular-momentum major mergers.

In this {\em Letter}, we present a set of simulated clusters in which,
for the first time, we find (1) co-existence of CC and NCC systems,
(2) agreement with the observed thermodynamical and chemical
properties of the two classes of clusters, and (3) evidence of a
recent transition between CC and NCC in the evolution of individual
objects. As we will discuss, this result is achieved thanks to the
combined action of a new AGN feedback model
\citep{steinborn.etal.2015} and an improved smoothed-particle-hydrodynamics (SPH) scheme
\citep{beck.etal.2015}.  

\section{Simulations}\label{sec:sim} 

We provide a short description of the simulations analyzed here, while
we refer to a forthcoming paper (Planelles et al. 2015, in preparation) for
a more thorough description. The simulations have been generated from
the same set of initial conditions originally presented by
\cite{bonafede.etal.11}. They correspond to zoomed-in initial
conditions of 29 Lagrangian regions selected around 24 massive
clusters with $M_{500}$ between 5 and 20 $\times 10^{14}$ \minv\  and 5 poorer
clusters with $M_{\rm 500}$ in the range 0.7--3$\times 10^{14}$\minv.
The simulations are carried out with the GADGET-3 code
\citep{springel2005} and include an upgraded version of the SPH
scheme, as described in \citet{beck.etal.2015}.  These developments
consist of (a) a higher-order (Wendland C$^4$ instead of the
standard B-spline) interpolating kernel to better describe
discontinuities and reduce clumpiness instability, (b) a
time-dependent artificial viscosity term to capture shocks and minimize viscosity away
from shock regions, (c) an artificial conduction term that
largely improves the SPH capability of following gas-dynamical
instabilities and mixing processes. 
With respect to
the physical thermal conduction, it similarly promotes the transport
of heat, 
since it is
applied to contact discontinuities in internal energy, but, differently, it does not depend on 
thermodynamical properties of the gas.  To assess the performances of this improved SPH
implementation, several standard hydrodynamical tests, such as weak
and strong shocks, shear flows, gas mixing, and self-gravitating gas,
have been performed in \cite{beck.etal.2015}.

The physical processes included in our simulations can be described as
follows.  Metallicity-dependent radiative cooling and the effect of a
uniform time-dependent UV background are considered as in
Planelles et al. (2015, see also \citealt{wiersma.etal.2009}). A
sub-resolution model for star formation from a multi-phase
interstellar medium is implemented as in
\cite{springel.hernquist.2003}. Kinetic feedback driven by supernovae (SN) is
accounted in the form of galactic winds with a velocity of $v_w \sim
350$ km s$^{-1}$. Metal production from SN-II, SN-Ia and
asymptotic-giant-branch (AGB) stars follows the original receipt by
Tornatore et al. (2007, see also \citealt{planelles.etal.2014}).
The AGN feedback is modeled with the implementations recently
presented in \cite{steinborn.etal.2015}.  The released thermal energy
accounts for contributions by both mechanical outflows and radiation,
separately computed in the code.  Eddington-limited gas accretion onto
BHs is computed by multiplying the Bondi rate by a boost factor
$\alpha$. In this {\em Letter}, we consider only cold accretion
($\alpha_{\rm cold}=100$ and $\alpha_{\rm hot}=0$). 
We verified that the results hold when hot accretion is
included with $\alpha_{\rm hot}=10$.  
%
For the feedback energy, we assume variable efficiencies
for the radiation and the mechanical outflow (Equations 19 and 20 of
\citealt{steinborn.etal.2015}) that depend on the BH mass and on the
accretion rate. Finally, we fix $\epsilon_f=0.05$ as the efficiency of
coupling the radiated BH energy to the ICM.

We emphasize that  the parameters defining the modification of SPH and the
AGN feedback model have been tuned to overcome the limitations of
standard SPH in passing classic hydrodynamical tests and to reproduce
the observational scaling relation between BH mass and stellar mass of
the host galaxies, including the brightest cluster galaxies
\citep{mcconnell&ma}. No attempt has been pursued to reproduce any of
the observational properties of the ICM. The existing match 
between other simulated and observed stellar and hot gas properties will be presented 
in future papers.

 The dark matter mass resolution is equal to $8.3 \times 10^8 h^{-1} M_{\odot}$ 
and the $z=0$ Plummer-equivalent softening length for the gravitational
 force is $\epsilon=3.75 h^{-1}$ kpc. In our sample, this value is always inferior 
 than $0.01 \times R_{500}$.
 
\section {Results}\label{sec:results} 

Several approaches have been used in the literature to classify
clusters as CC and NCC systems based, e.g., on the central gradient of
temperature or entropy, on the central entropy level, on the cooling time,
on the mass deposition rate inferred from X-ray spectroscopy, on the
cuspiness of the central gas density profile \citep[comparisons among
  these criteria are presented
  in][]{cavagnolo.etal.09,hudson.etal.2010,leccardi.etal.2010,mcdonald.etal.2013,
  pascut_ponman}.  In order to assign a degree of ``cool-coreness'' to
our simulated clusters, we jointly use the following two criteria
which measure the shape and level of the entropy profiles in the
central regions and are extensively applied to observational data.

Following \citet[][see also Leccardi et al. 2010]{rossetti.etal.2011},
we base the first criterion on the computation of the pseudo entropy:
\begin{equation}
\sigma=\frac{(T_{\rm IN}/T_{\rm OUT}) }{({\rm EM}_{\rm IN}/{\rm EM}_{\rm OUT})^{1/3}},
 \end{equation}
where the spectroscopic-like temperature \citep{mazzotta.etal.2004},
$T$, and the emission measures, EM, are computed in the IN region,
$r<0.05 \times R_{180}$, and in the OUT region, $0.05 \times
R_{180}<r<0.15\times R_{180}$. We define CC those clusters with
$\sigma<0.55$.

In addition, we request the value of the central entropy to be
$K_0<60$ keV cm$^2$.  We model the entropy profile as
\begin{equation}
K=K_0 + K_{100} \times (R/R_{100k})^{\alpha},
\end{equation}
where we treat as free parameters the core entropy, $K_0$, the entropy
at $R_{\rm 100k}(=100$ kpc), $K_{100}$, and the slope, $\alpha$,
\citep{cavagnolo.etal.09}. For each simulated cluster the entropy
profile is measured in linearly equispaced spherical shells, with the
innermost radius being that of the sphere containing at least 100 gas
particles. In all cases, this radius is greater than $5\times \epsilon$. 
We use $R_{500}$ as the outermost radius out to which
we perform the fit. Three clusters with $\sigma<0.55$ have steadily
declining entropy profiles.  In these cases, $K_0$ is not
constrained. We, thus, consider the overall profiles and keep in the
CC sample the two objects with $\alpha \sim 1$ and exclude that with
$\alpha<0.5$.

\begin{figure*}[ht!]
\centering
\includegraphics[width=0.45\textwidth]{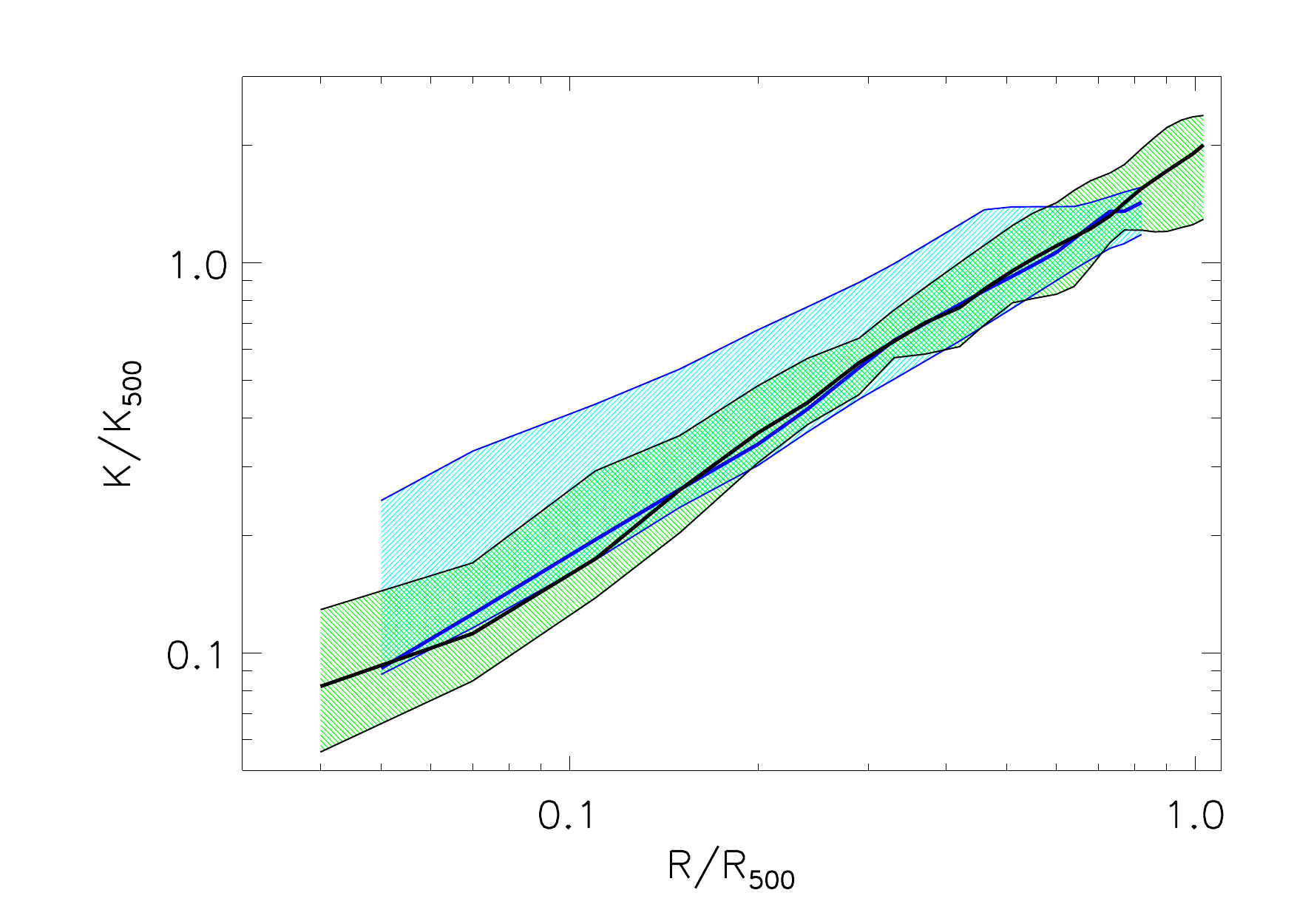}
\includegraphics[width=0.45\textwidth]{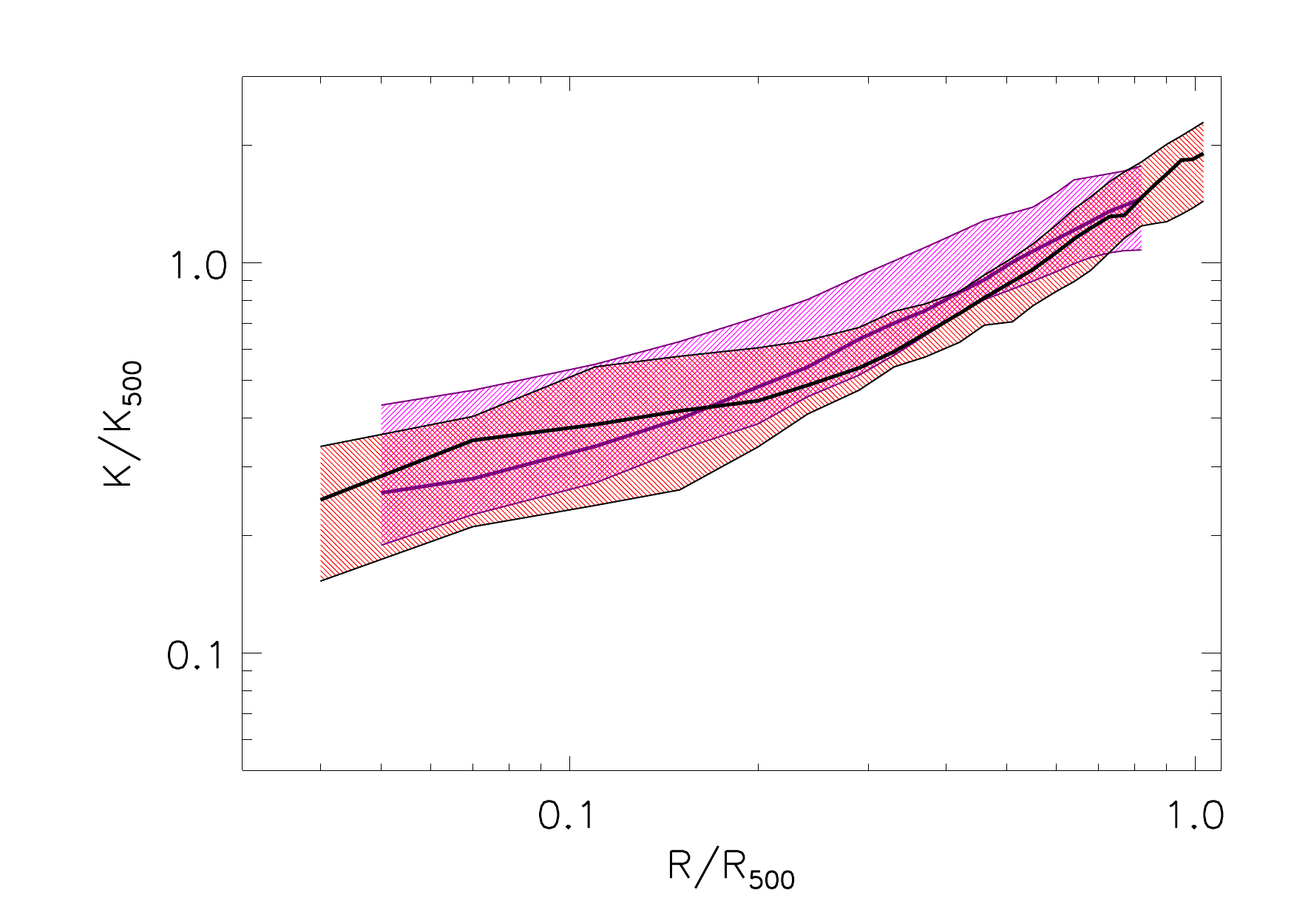}
\includegraphics[width=0.45\textwidth]{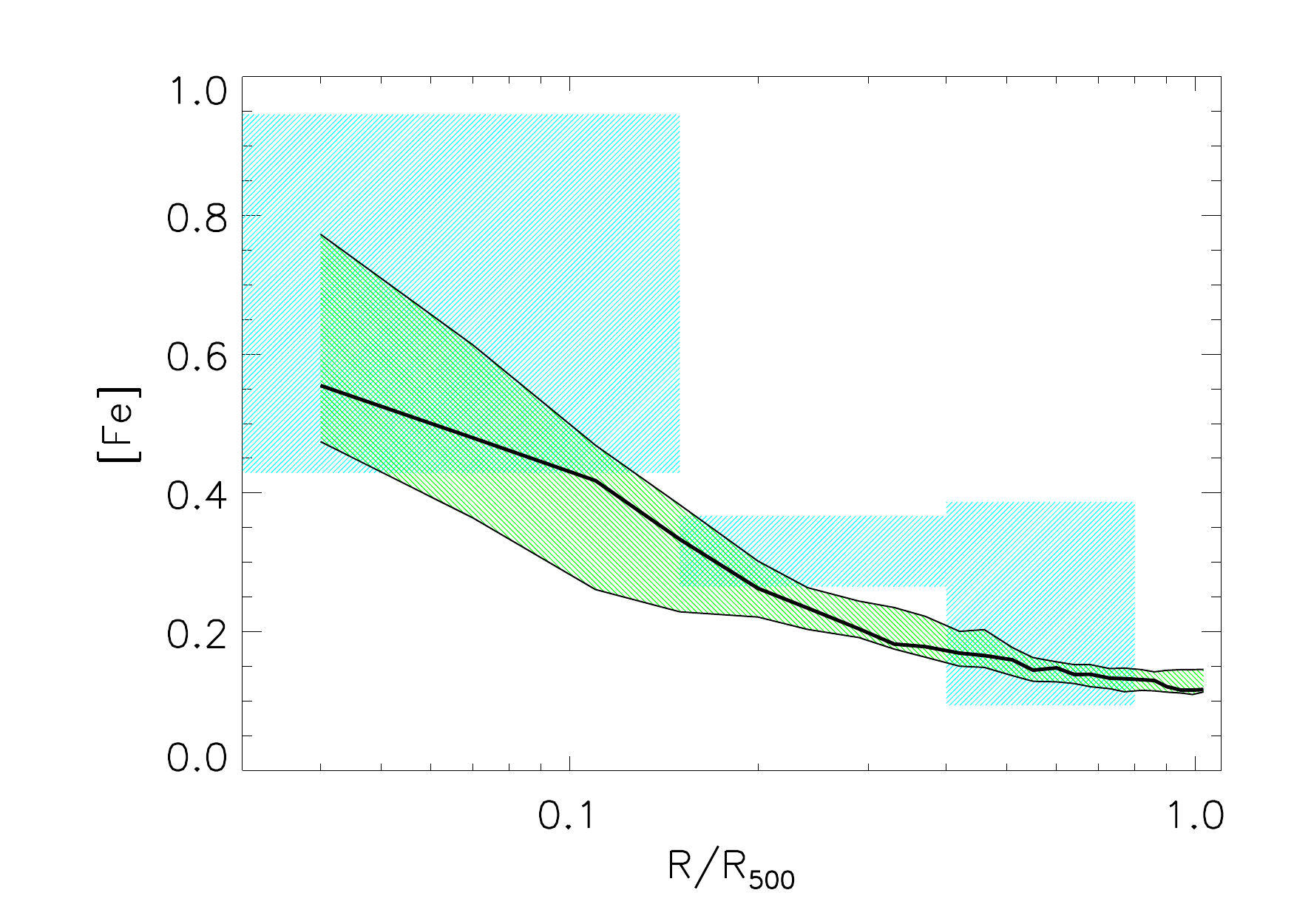}
\includegraphics[width=0.45\textwidth]{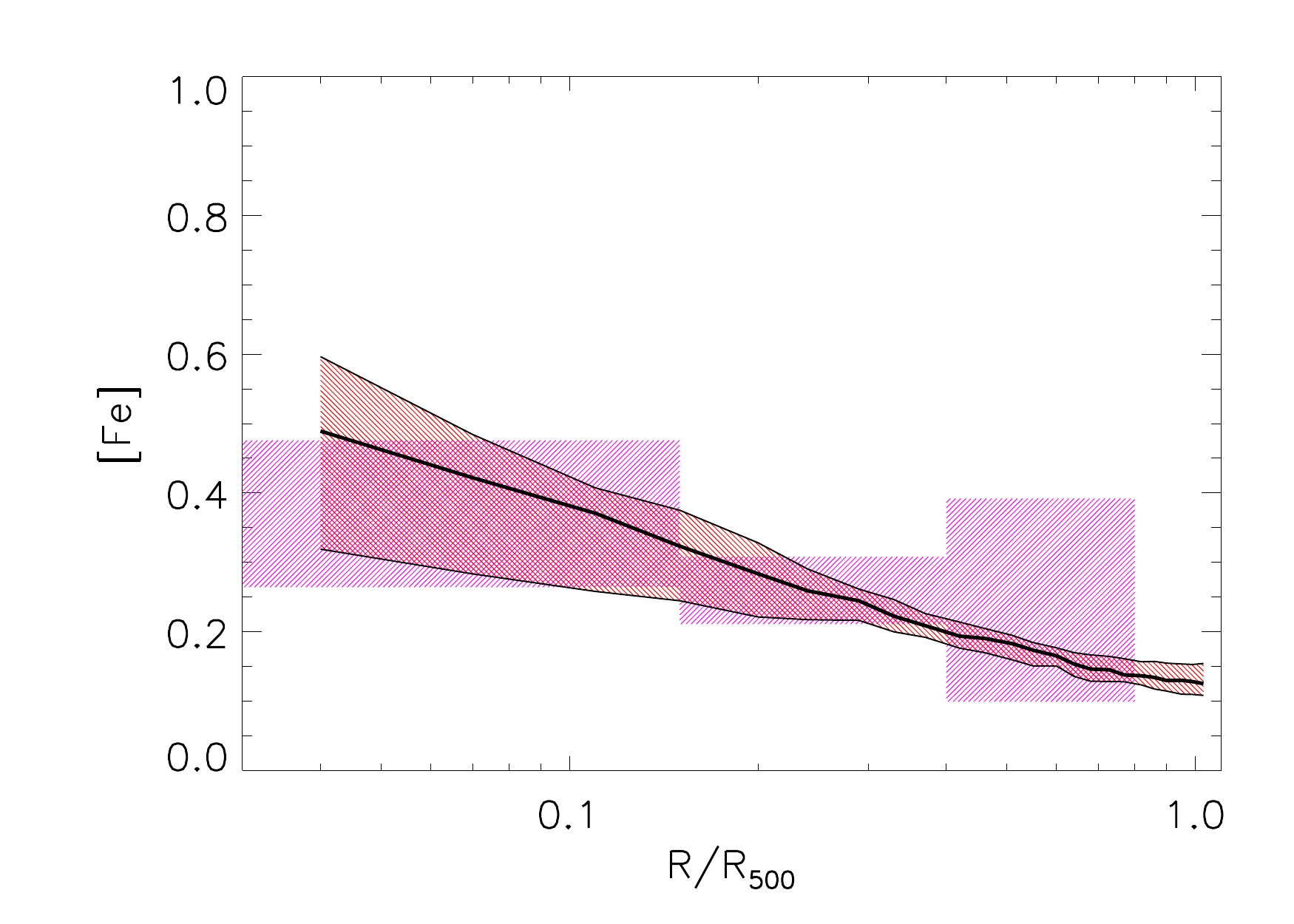}
\caption{ Simulated $z=0$ entropy (top panels) and Iron abundance (bottom
  panels) profiles compared with observations by
  \cite{pratt.etal.2010} and by \cite{ettori.etal.2015},
  respectively. For the simulated CC (left panels) and NCC
  (right panels) profiles: the medians are in black and the 16$^{th}$ and the 84$^{th}$
  percentiles are delimited by shaded regions. The observed profiles are 
  in cyan (CC) and magenta (NCC). Metallicity profiles are expressed in solar units using
  \cite{anders_grevesse}.}
\label{fig:prof}
\end{figure*}

The distribution of $\sigma$ and $K_0$ is continuous and no clear 
bimodality is present, in contrast with recent results by \cite{hahn.etal.2015}. 
Considering the aforementioned observational thresholds, we 
find that 38\% (11/29) of our simulated clusters at $z=0$
are classified as CC. This fraction is quite close to the 35\% value
reported by \cite{eckert.etal.2011} from a similar mass distributed sample, 
HIGFLUGS, after accounting for the sample incompleteness and the X-ray selection bias.

\begin{figure*}[ht!]
\centering
\includegraphics[width=0.9\textwidth]{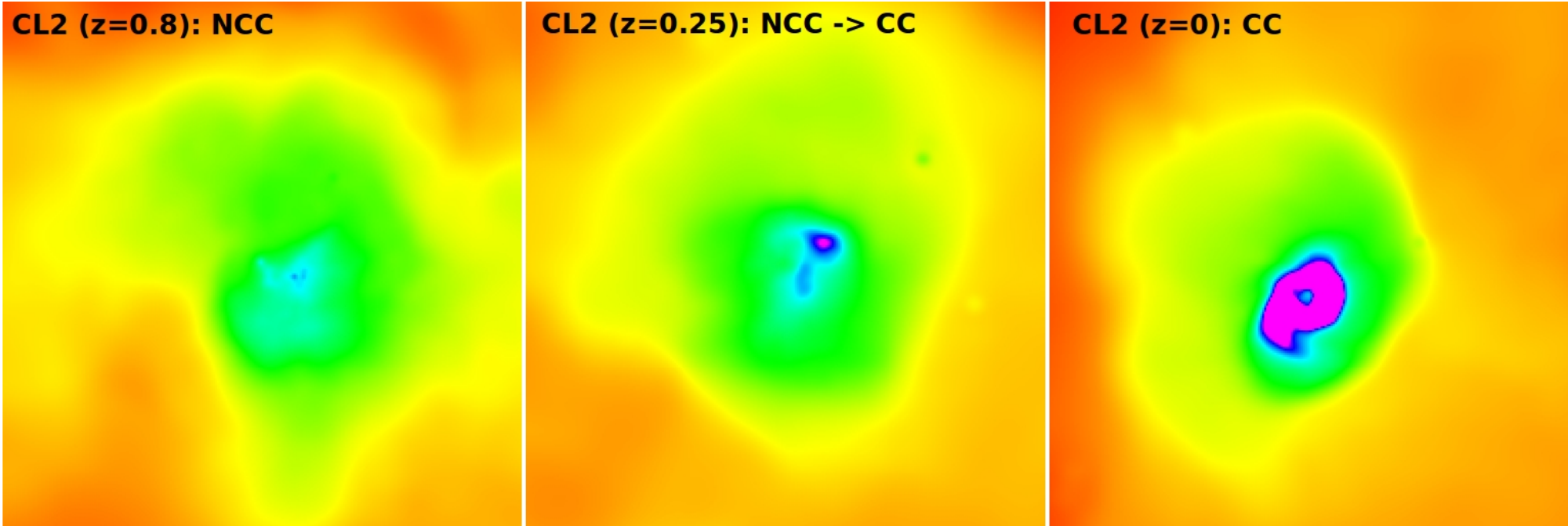}
\includegraphics[width=0.9\textwidth,trim=0 20 0 0,clip]{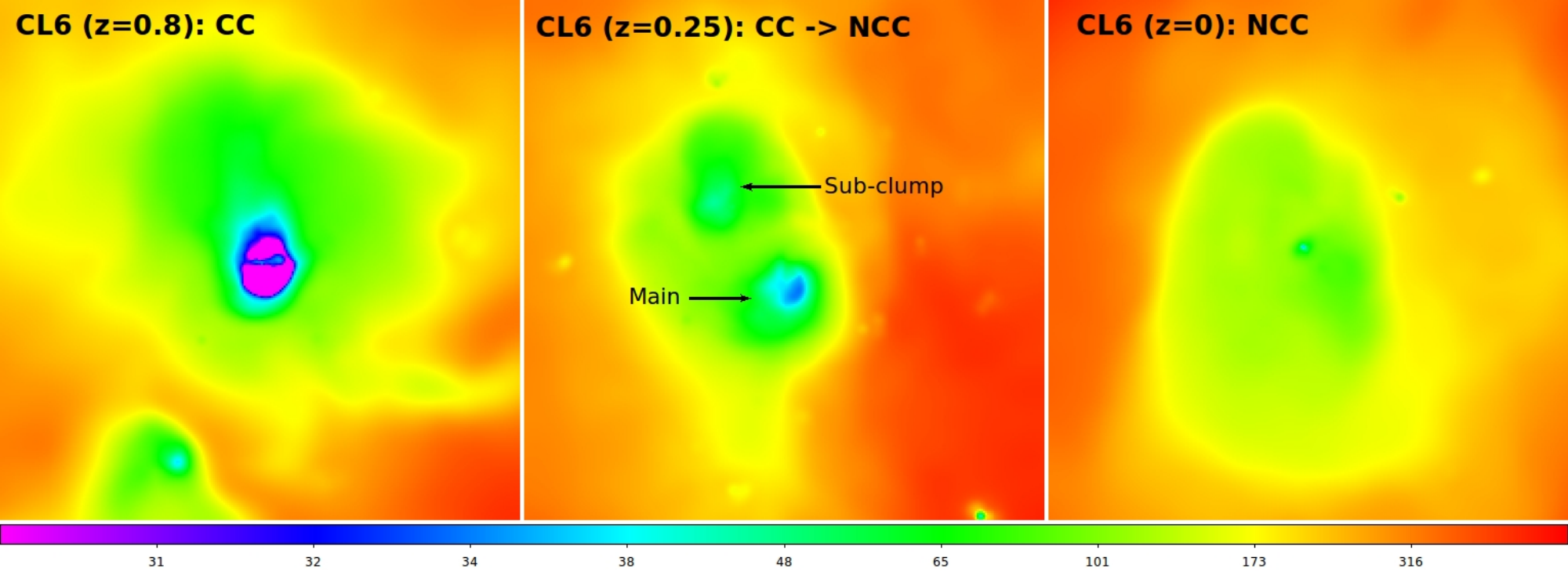}
\includegraphics[width=0.035\textwidth,angle=270,trim=10 0 0 0,clip]{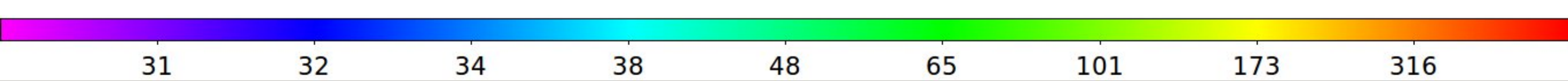}
\caption{Maps of pseudo entropy of two simulated clusters with masses $M_{500}=2.4 \times 10^{14}$\minv (upper panels)
  and $M_{500}=7.3 \times 10^{14}$\minv (lower panels) at $z=0$. 
  The size of the images is 1 Mpc and the line-of-sight integration is for 10 Mpc.
}
\label{fig:map}
\end{figure*}

In the upper panels of Figure \ref{fig:prof}, we compare the entropy
profiles of our simulated clusters with observational results obtained
by \citet[][]{pratt.etal.2010} from {\em XMM-Newton} observations of the
REXCESS clusters.  Our simulations now produce a population of CC
clusters, whose entropy profiles decline down to the innermost
regions. This result confirms that the energy extracted from the
central BHs is distributed in such a way to compensate for the radiative
losses of gas at low entropy and, therefore, to keep it in the hot
phase despite the fact that it formally had a short cooling time.  More generally,
we note that observed and simulated entropy profiles agree quite well
in normalization and slope, for both CC and NCC populations.

 \cite{dubois.etal.11} were also able to produce the correct 
CC entropy structure in a high-resolution simulation of a Virgo-like cluster by
including the effect of AGN feedback. However, in the same paper, these
authors claimed that this conclusion is spoiled as soon as the
effect of metallicity is included in the cooling function. This
confirms the fragility of a self-regulated AGN feedback. 
We, therefore, compare 
 the simulated mass-weighted
iron profiles to the metallicity of nearby clusters,
$z<0.2$, from \cite{ettori.etal.2015}\footnote{In their classification of CC and NCC, \citet{ettori.etal.2015} used a
  different definition of the IN and OUT regions in the
  pseudo-entropy calculation.  For their sample, we consider the
  limit $\sigma <0.6$ that corresponds to our adopted threshold
  ($\sigma <0.55$) according to \citet{baldi.etal.2012}.}, in the bottom panels of Figure \ref{fig:prof}.  
  In keeping
with observational results, simulated CC clusters are characterized
by a rather high enrichment level in the central regions while NCC
objects have a shallower peak.  This agreement suggests that our
simulations provide an 
 {\em effective} description of the
mixing related to gas-dynamical processes acting during the cluster
formation and of the rise of enriched gas caused by feedback
processes: the same processes leading to the diversity of entropy
profiles are also responsible for the diversity of the profiles of
metal abundance.

From an object-by-object investigation of our simulated clusters, we
recognize a variety of situations for the establishment of a CC and
for the transformation between CC and NCC, which can take place over a
quite a short time and even at low redshift. As an example, in Figure
\ref{fig:map} we show the evolutionary sequences of pseudo-entropy maps
from $z=0.8$ to $z=0$ for two objects. Following an observational
approach, we create maps of spectroscopic-like temperature, ${\cal
  M}_{\rm TSL}$, and of X-ray soft band ([0.5-2] keV) emissivity,
${\cal M}_{\rm X}$, which we then combine as ${\cal M}_{\rm TSL}/{\cal
  M}_{\rm X}^{1/3}$ to build maps of pseudo entropy
\citep{finoguenov.etal.2010}. The upper panel shows the case of an NCC
cluster at $z=0.8$ that slowly and continuously develops a CC structure.  On the
contrary, the lower panel corresponds to a system with a low-entropy
core at high redshift. The core is still in place at $z=0.25$ and it
is eventually destroyed by a merger between the main halo and a
sub-clump at $z\sim 0.1$. Remarkably, this last case of CC remnant is
not an isolated event: almost half of NCC clusters at $z=0$ were CC at
$z=1$ \citep{rossetti.etal.2011}.

These results vary somewhat from previous claims, also
based on simulations, which stated that CCs can only be destroyed at high redshift
through major mergers, while being much more resilient against mergers
at low redshift \citep[e.g.][]{burns.etal.2008,poole.etal.2008}.  A
close comparison of our results with these previous analyses is not
straightforward due to substantial differences in both the
hydrodynamic schemes and the implementation of
feedback. \cite{poole.etal.2008} carried out non-cosmological
simulations of isolated two-body mergers using a more standard
implementation of SPH, while \cite{burns.etal.2008} used a Eulerian
code with a cosmological set-up.  Neither one of these two analyses
included the effect of AGN feedback and, for this reason, the strength
of their cores is probably affected by overcooling. In order to
surpass this limitation, \cite{burns.etal.2008} truncated star
formation below a certain redshift, while \cite{poole.etal.2008}
excluded the central 40 kpc in their analysis. Both strategies,
however, do not directly attack the main problem of the delicate
balance between cooling and heating.  This balance, essential for the
creation of a low-entropy core, needs to be regulated since early
times and can be achieved in our simulations only by including AGN
feedback.

\section{Discussion and conclusions}

It has long been known that ICM evolution purely driven by gravitational
processes leads to self-similar entropy profiles that scale as a
power-law of radius, $K(r)\propto r^\alpha$ with $\alpha \sim 1$--1.1
in the cluster outskirts \citep{voit.etal.2005,nagai.etal.2007}.
Non-radiative simulations, however, differ in their behavior in the
central regions.  A flat core profile is found in grid-based codes
\citep{frenk.etal.1999} and in recent improved versions of SPH
\citep[][and references
  therein]{biffi_valdarnini,sembolini.etal.2015}.  Standard SPH
simulations, instead, produce entropy profiles that are declining
toward the center.

In addition to mixing, radiative processes have the counterintuitive
effect of also raising entropy in core regions
\citep[e.g.][]{borgani&kravtsov}. If not counteracted by some energy
feedback process, radiative cooling selectively removes gas with low
entropy and short cooling time from the hot-X-ray emitting phase, thus
fuelling star formation \citep[e.g.][]{li.etal.2012} and leaving in
the ICM only gas with a relatively higher entropy.

The AGN feedback plays a key role in explaining our results. Considering
the new SPH set-up but excluding AGN feedback, we find that
overcooling leads to a thermal structure of central cluster regions
whose entropy profiles behave as in NCC clusters (top panel of
Figure~3).  At the same time, the excess of star formation produces an
exceedingly high level of ICM metal enrichment, with heavy elements
mostly locked around galaxies. As a consequence, simulated clusters
without AGNs always display spikes of central metallicity, an enrichment
pattern typical of CC clusters (bottom panel of Figure~3).  As
observations suggest, the AGN feedback should have the threefold
effect of (1) compensating radiative losses so as to prevent gas with
a short cooling time from dropping out of the ICM, (2) reducing star
formation in massive halos at low redshift, (3) raising metals
away from star-forming regions
 \citep[e.g.][]{sijacki.etal.2007,fabjan.etal.2011,mccarthy.etal.2011}.

However, attempts to create realistic thermal structures of CC in
simulated clusters so far have failed, even when including AGN
feedback, which is otherwise successful in reproducing X-ray scaling
relations   \citep[e.g.][]{puchwein.etal.2008,planelles.etal.2014} and their scatter \citep{lebrun.etal.2014}.  
The encouraging success of the simulations presented here in producing 
thermal and chemo-dynamical structure of CC 
in agreement with observations 
is related to the
combined action of the AGN feedback model {\em and} of the artificial
thermal diffusion.

\begin{figure} 
\centering
\includegraphics[width=0.45\textwidth]{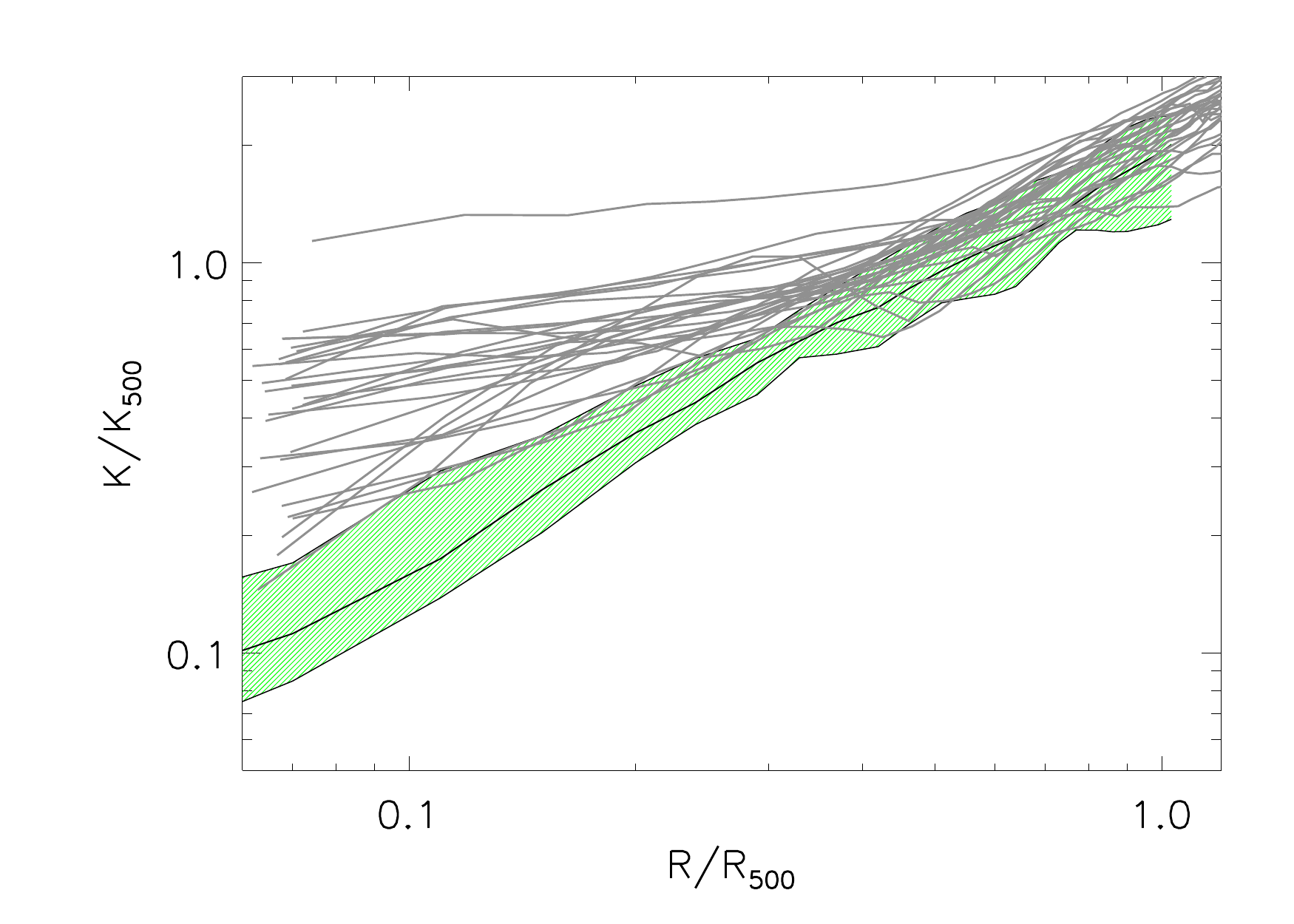}
\includegraphics[width=0.45\textwidth]{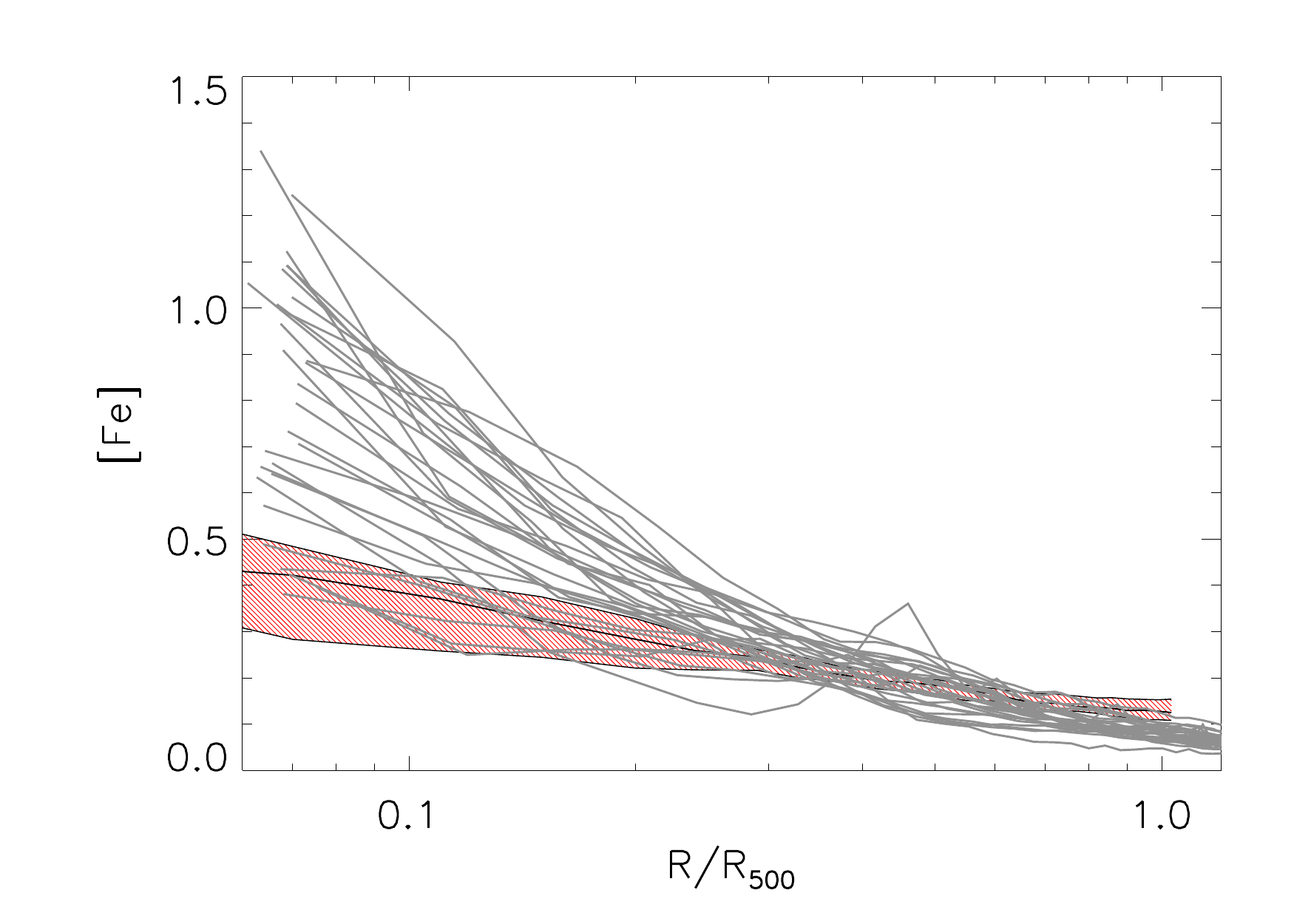}
\caption{The effect of AGN feedback on entropy (upper panel) and iron
  abundance (lower panel) profiles for $z=0$ clusters. In each panel
  the grey lines  are the profiles of all clusters simulated without AGN.
  For comparison, we report the shaded area of the CC entropy (from Fig. 1, top-left panel)
  and the shaded area of the NCC metallicity (from Fig. 1, bottom- right panel). }
\label{fig:prof_noAGN}
\end{figure}

\begin{figure}
\centering
\includegraphics[width=0.5\textwidth]{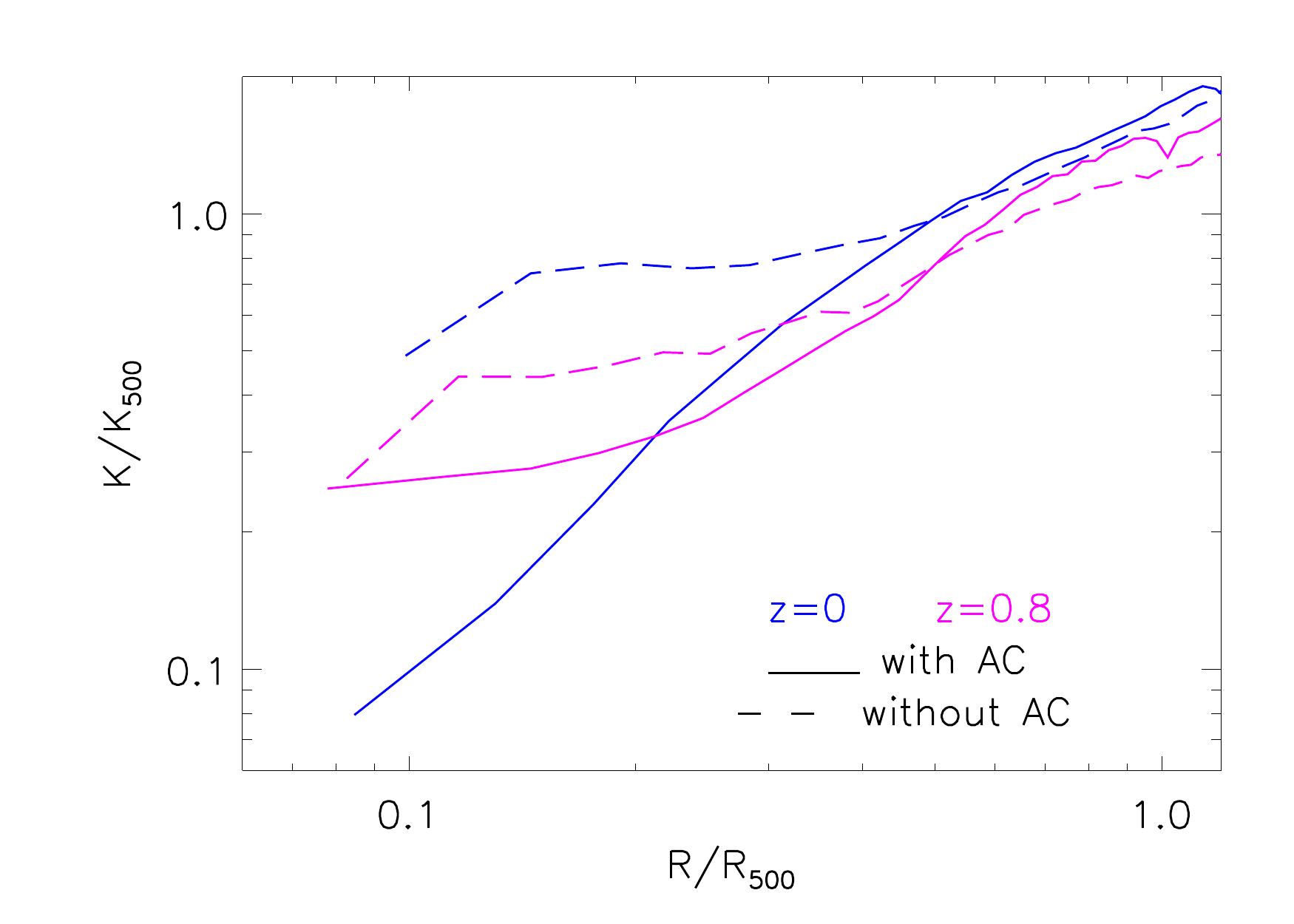}
\caption{Effect of artificial conduction on the entropy profiles of
  the cluster shown in the upper panel of Fig. \protect\ref{fig:map}
  at two instances: $z=0.8$ (magenta curves) and $z=0$ (blue curves).
  The simulation has been carried out with and without artificial
  conduction (solid and dashed lines, respectively), while leaving all
  the other parameters unchanged. The cluster simulated with
  artificial conduction was classified NCC and CC at the two epochs.}
\label{fig:test_nocond}
\end{figure}

Figure~\ref{fig:test_nocond} compares the entropy profiles of the
cluster shown in the upper panels of Figure \ref{fig:map} with those
of the same object simulated without artificial conduction but
includes other recent implementations such as the new kernel and the
artificial viscosity. 
By excluding artificial conduction, the entropy profiles at $z=0.8$ and $z=0$ are similar to
each other with a large isentropic core, characteristics of a strongly
NCC object.  This contrast with the previous finding showing a
transition from an NCC to a CC situation (Figure~\ref{fig:map}).
Besides confirming the results reported in previous works based on
standard SPH \citep{sijacki.etal.2008,planelles.etal.2014}, this
outcome highlights the role of artificial conduction in providing an
effective description of the processes of mixing that lead to the
redistribution of AGN feedback energy in central regions.

\vspace{0.3cm}
 In summary, we showed that our simulations produce a population of
clusters that have a CC reasonable structure, 
both from a thermo- and
chemo-dynamical point of view.  Furthermore, once this result is
achieved, the simulations also naturally produce a mix of CC and NCC
clusters, which are similar to that observed.  We trace the reason for
the success of our simulations to the combined action of artificial
conduction, introduced to overcome numerical limitations of standard
SPH, and of AGN feedback, introduced to regulate star formation in
massive halos.

As a concluding word of caution, we would like to emphasize that the
success of the simulations presented here does not imply that we are
self-consistently describing the details of all of the physical processes
that lead to the creation of cool cores: inflation of bubbles of
high-entropy gas from the shocks of sub-relativistic jets launched by
the AGN, gas circulation and turbulence triggered by the buoyancy of
these bubbles, possible effects of magnetic fields in stabilizing
them, thermal conduction, and cosmic rays. These are all effects for which
we have circumstantial evidence from observations, but that are not
included in our simulations. Still, our results demonstrate that our
description of AGN feedback and artificial conduction provides a
realistic {\em effective} description of the above physical processes,
without requiring any specific fine tuning.

\acknowledgments

We are greatly indebted to Volker Springel for the access to
the GADGET3 code; to D. Fabjan, V. Fiorenzo, M. Petkova, and
L. Tornatore for the simulation set-up; and
 to the referee, D. Eckert, S. Ettori, A. Evrard, M. Gaspari, S. Molendi,
P. Monaco, P. Tozzi, M. Voit for useful discussions. We acknowledge financial support from
 PIIF-GA- 2013-627474, NSF
AST-1210973, PRIN-MIUR 201278X4FL, PRIN-INAF 2012 "The Universe in a
Box: Multi-scale Simulations of Cosmic Structures", the INFN INDARK
grant, ``Consorzio per la Fisica'' of Trieste, DFC Cluster of
Excellence `Universe', DFC Research Unit 1254, CONICET-Argentina, FonCyT.
Simulations are carried out using Flux HCP Cluster at the University
of Michigan, Galileo at CINECA (Italy), with CPU time assigned through ISCRA proposals and an
agreement with the University of Trieste, and PICO at CINECA though our expression of interest.

%

\end{document}